\documentclass[twocolumn]{aastex61}
\pdfoutput=1

\newcommand{\etal}{et al.}
 
\submitjournal{ApJ}

\shortauthors{Chianese \etal}

\begin{document}

\title{Use of ANTARES and IceCube data to constrain a single power-law neutrino flux}

\correspondingauthor{Marco Chianese}
\email{chianese@na.infn.it}

\author{Marco Chianese}
\affiliation{Dipartimento di Fisica {\it Ettore Pancini}, Universit\`a di Napoli Federico II, Complesso Univ. Monte S. Angelo, I-80126 Napoli, Italy}
\affiliation{INFN, Sezione di Napoli, Complesso Univ. Monte S. Angelo, I-80126 Napoli, Italy}

\author{Rosa Mele}
\affiliation{Dipartimento di Fisica {\it Ettore Pancini}, Universit\`a di Napoli Federico II, Complesso Univ. Monte S. Angelo, I-80126 Napoli, Italy}
\affiliation{INFN, Sezione di Napoli, Complesso Univ. Monte S. Angelo, I-80126 Napoli, Italy}

\author{Gennaro Miele}
\affiliation{Dipartimento di Fisica {\it Ettore Pancini}, Universit\`a di Napoli Federico II, Complesso Univ. Monte S. Angelo, I-80126 Napoli, Italy}
\affiliation{INFN, Sezione di Napoli, Complesso Univ. Monte S. Angelo, I-80126 Napoli, Italy}

\author{Pasquale Migliozzi}
\affiliation{INFN, Sezione di Napoli, Complesso Univ. Monte S. Angelo, I-80126 Napoli, Italy}

\author{Stefano Morisi}
\affiliation{Dipartimento di Fisica {\it Ettore Pancini}, Universit\`a di Napoli Federico II, Complesso Univ. Monte S. Angelo, I-80126 Napoli, Italy}
\affiliation{INFN, Sezione di Napoli, Complesso Univ. Monte S. Angelo, I-80126 Napoli, Italy}

\begin{abstract}
We perform the first statistical combined analysis of the diffuse neutrino flux observed by ANTARES (nine-year) and IceCube (six-year) by assuming a single astrophysical power-law flux. The combined analysis reduces by a few percent the best-fit values for the flux normalization and the spectral index. Both data samples show an excess in the same energy range (40--200~TeV), suggesting the presence of a second component. We perform a goodness-of-fit test to scrutinize the null assumption of a single power-law, scanning different values for the spectral index. The addition of the ANTARES data reduces the $p$-value by a factor 2$\div$3. In particular, a single power-law component in the neutrino flux with the spectral index deduced by the six-year up-going muon neutrinos of IceCube is disfavored with a $p$-value smaller than $10^{-2}$. 
\end{abstract}


\section{Introduction}

Neutrino telescopes (NTs) can play a crucial role in unveiling the physical properties of astrophysical high-energy emitters such as supernovae, active galactic nuclei, gamma-ray bursts, etc.,  or in the possible discovery of genuine new astrophysical sources. At the same time, NTs could also make important contributions to a deeper understanding of neutrino physics. This can occur, for example, with the discovery of the correct mass ordering, or via the observation of new physics in neutrino oscillation phenomena such as sterile neutrinos~\citep{Aartsen:2017bap} or non-standard interactions~\citep{Day:2016shw}, etc. More recently, after the IceCube first observation of astrophysical high-energy neutrinos, the possibility to indirectly detect heavy dark matter through its emission in neutrinos has been extensively discussed in literature~\citep{Feldstein:2013kka}.

In this paper, we focus on the recent preliminary data presented by ANTARES (A;~\citealt{proceedingA}) and IceCube (IC;~\citealt{proceedingIC}) Collaborations at the ICRC 2017 Conference, and we perform a combined analysis of both observations. According to~\citet{proceedingA}, the ANTARES Collaboration has reported in the nine-year data sample the observation of an excess of neutrinos over the atmospheric background for energies above 20~TeV. In particular, $14$ shower events have been detected from 2007 to 2015, while $10.5\pm4.0$ are compatible with the expected atmospheric background that is dominated by penetrating muons. Moreover, once the track events are also taken into account, the observed excess in ANTARES has a $p$-value of 0.15 (below $2\sigma$;~\citealt{proceedingA}). Even though such an excess is still compatible with the background within its uncertainties, it is not in disagreement with the expectations from the astrophysical power-law flux observed by IceCube~\citep{Kopper:2015vzf}.

After six years of data-taking, the IceCube Collaboration has collected a total of 82 events (track + shower) with a neutrino interaction vertex that is located inside the detector~\citep{proceedingIC}. This data sample is called as ``High-Energy Starting Events'' (HESE). The analysis performed by considering neutrinos that have energy above 60~TeV provides a best-fit power-law with spectral index equal to $\gamma^{\rm 6yr}_{\rm IC}=2.92^{+0.29}_{-0.33}$. Such a best-fit spectral index is larger than the one obtained by the four-year HESE data ($\gamma^{\rm 4yr}_{\rm IC}=2.58\pm0.25$;~\citealt{Kopper:2015vzf}). This is due to the fact that during the last two years, no neutrinos with energy larger than 200~TeV have been observed.

In general, one would expect a hard power-law behavior for the neutrino flux. For instance,~\citet{Waxman:1998yy} predicted a cosmic neutrino flux proportional to $E^{-2.0}$, according to the standard Fermi acceleration mechanism at shock fronts~\citep{Fermi:1949ee}. For this reason, anomalous large values for the spectral index such as $\gamma^{\rm 4yr}_{\rm IC}$ and $\gamma^{\rm 6yr}_{\rm IC}$ have suggested the presence of an additional component dominating at lower energies ($E_\nu \leq 200$~TeV), on top of a single unbroken power-law flux (see~\citealt{Chen:2014gxa,Palladino:2016zoe,Palladino:2016xsy,Vincent:2016nut,,Anchordoqui:2016ewn,proceedingIC} for studies of a neutrino flux with two power-laws and~\citealt{Kimura:2014jba,Murase:2015xka,Senno:2015tsn} for viable hidden astrophysical candidates). The two-component scenario is also strongly motivated by the analyses performed on the up-going muon neutrinos~\citep{Aartsen:2016xlq}. Indeed, the recent six-year data sample points toward a harder power-law with spectral index of $2.13\pm0.13$~\citep{Aartsen:2016xlq}. Such a value is in a~3.3$\sigma$ tension with the combined analysis of different IceCube data samples~\citep{Aartsen:2015knd}.

Once a hard power-law is considered according to the analyses of up-going muon neutrinos~\citep{Aartsen:2016xlq}, a low-energy excess in the range 40--200~TeV arises in different IceCube data samples~\citep{Aartsen:2014muf,Kopper:2015vzf}. The statistical characterization of such a $\sim2\sigma$ excess, with respect to a power-law with spectral index 2.0, was highlighted in~\citet{Chianese:2016opp} by analyzing the four-year HESE data~\citep{Kopper:2015vzf}. The local statistical significance of the excess increases up to $2.3\sigma$ in the case of two-year ``Medium-Energy Starting Events'' (MESE;~\citealt{Aartsen:2014muf}), as shown in~\citet{Chianese:2016kpu}.

\section{Analysis with a single power-law}

In this study, we perform a combined analysis of the diffuse neutrino flux observed by ANTARES (nine-year showers;~\citealt{proceedingA}) and by IceCube (six-year HESE showers and tracks;~\citealt{proceedingIC}), assuming a single astrophysical component parametrized in terms of the following unbroken power-law per neutrino flavor
\begin{equation}
\frac{{\rm d}\Phi_{\rm astro}}{{\rm d}E_\nu{\rm d}\Omega} = \Phi_{\rm astro}^0 \left(\frac{E_\nu}{100~{\rm TeV}}\right)^{-\gamma_{\rm astro}}\,.
\label{eq:flux}
\end{equation}
We consider an equal flavor composition at the Earth, as expected for standard astrophysical sources due to neutrino oscillations, and an isotropic flux in angular coordinates. In particular, we show how much the fit on the parameters $\left(\Phi_{\rm astro}^0,\gamma_{\rm astro}\right)$ with the IceCube data changes by including also the ANTARES observations in the analysis. Moreover, we also statistically characterize the low-energy excess as a function of the spectral index $\gamma_{\rm astro}$ by considering both ANTARES and IceCube measurements.

In addition to the astrophysical flux described by Eq.~(\ref{eq:flux}), we consider the conventional atmospheric background that consists of penetrating muons and neutrinos produced by the $\pi/K$ decays in the atmosphere~\citep{Honda:2006qj}. On the other hand, we do not take into account the prompt atmospheric background (neutrinos produced by the decays of charmed mesons;~\citealt{Enberg:2008te}), according to the IceCube conclusions contained in Ref.~\citep{Aartsen:2013eka,Aartsen:2014muf,Aartsen:2016xlq}.\footnote{In case of ANTARES data, we have subtracted the prompt component to the total background reported in Figure~1 of Ref.~\citep{proceedingA} through the neutrino effective area deduced by the same plot.}

In the present analysis, the combined fit is performed by maximizing the binned multi-Poisson likelihoods $\mathcal{L^{\rm A,\, IC}}$, whit expressions that are given by~\citet{Baker:1983tu}
\begin{equation}
\ln \mathcal{L^{\rm A,\, IC}} = \sum_i\left[n_i^{\rm A,\, IC} - N_i^{\rm A,\, IC} + n_i^{\rm A,\, IC} \ln\left(\frac{N_i^{\rm A,\, IC}}{n_i^{\rm A,\, IC}}\right)\right]\,,
\end{equation}
where the quantity $N_i$ is the expected number of events that is compared with the observed number of neutrinos $n_i$, once the atmospheric background events (conventional neutrinos and penetrating muons only) have been subtracted in each energy bin $i$. Moreover, the index $i$ runs over the energy bins of the two experiments. According to the cuts considered by the two Collaborations, we consider only neutrino events with $E_\nu\geq20$~TeV for ANTARES and $E_\nu\geq60$~TeV for IceCube. The expected number of events in the energy bin $\left[E_i,E_{i+1}\right]$ is then given by
\begin{equation}
N_i^{\rm A,\, IC} = 4\pi \Delta t^{\rm A,\, IC}\int_{4\pi}{\rm d}\Omega \int_{E_i}^{E_{i+1}}{\rm d}E_\nu \frac{{\rm d}\Phi}{{\rm d}E_\nu{\rm d}\Omega} A^{\rm A,\, IC}_{\rm eff}\left(E_\nu\right)\,,
\end{equation}
where $\Delta t^{\rm A,\, IC}$ is the exposure time ($\Delta t^{\rm A}=2450$~days and $\Delta t^{\rm IC}=2078$~days), whereas $A^{\rm A,\, IC}_{\rm eff}\left(E_\nu\right)$ is the effective area summed over the three neutrino flavors. The ANTARES effective area has been obtained by using the two cosmic neutrino spectra reported in Figure~1 of~\citet{proceedingA} assuming a constant effective area in each energy bin. The IceCube effective area, instead, has been taken from~\citet{Aartsen:2013jdh}.

\begin{figure}[t!]
\centering
\includegraphics[width=0.48\textwidth]{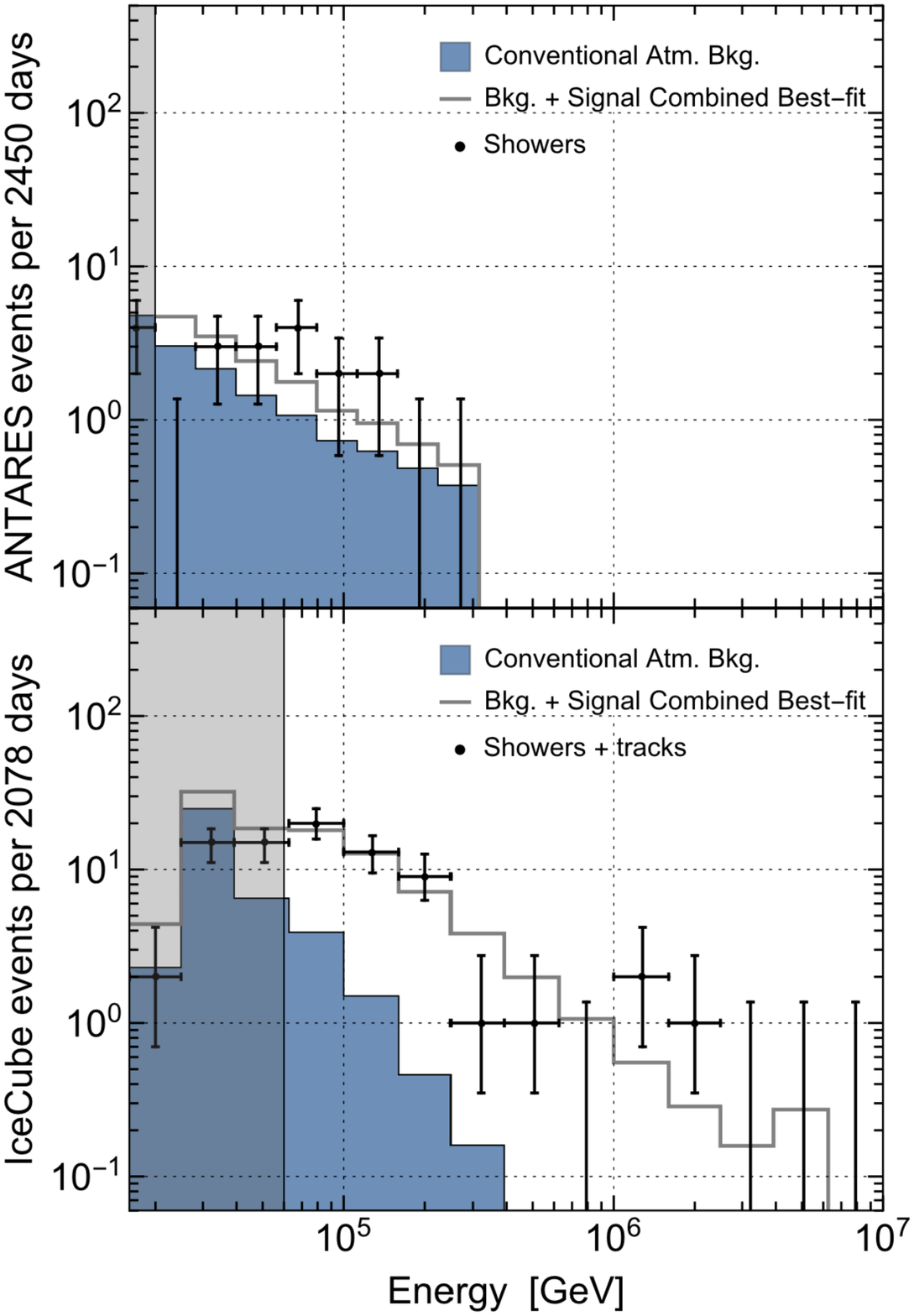}
\caption{\label{fig:best} Number of neutrino events as a function of the energy for ANTARES (upper panel) and IceCube (lower panel). In both plots, the blue area represents the conventional atmospheric background (neutrinos and penetrating muons), and the gray line corresponds to the sum of the background and the combined (IC+A) best-fit power-law with $\Phi_{\rm astro}^{0}=2.30\times10^{-18} \left({\rm GeV\,cm^2\,s\,sr}\right)^{-1}$ and $\gamma_{\rm astro}=2.85$.}
\end{figure}

\begin{figure}[t!]
\centering
\includegraphics[width=0.48\textwidth]{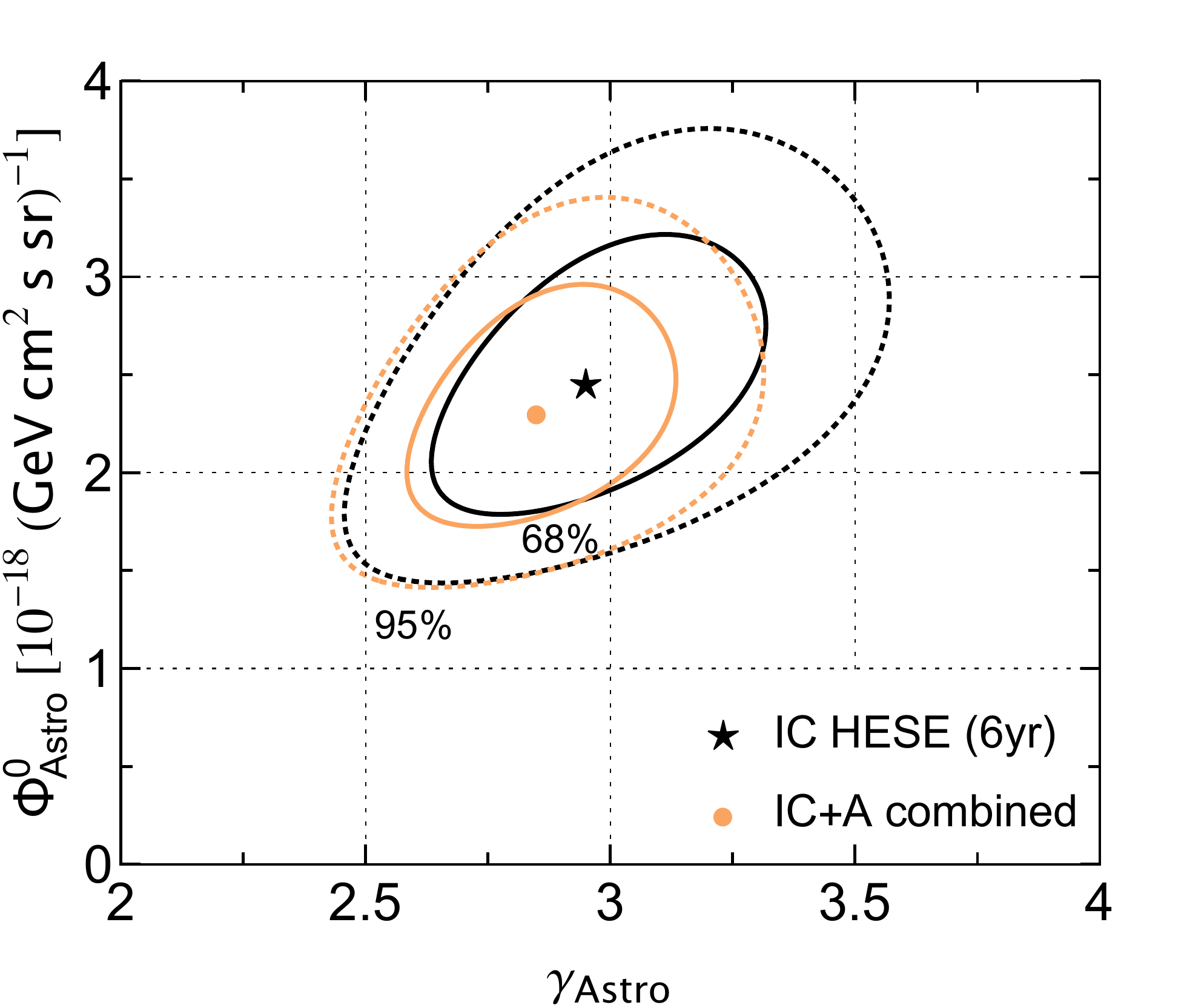}
\caption{\label{fig:bestpar}Contour plot of the likelihood $\mathcal{L}\left(\Phi_{\rm astro}^0,\gamma_{\rm astro}\right)$, in the case of the IceCube fit (black) and the combined IceCube+ANTARES one (ocher).}
\end{figure}

The combined fit (IC+A) is obtained by considering the product of the two likelihoods
\begin{equation}
\ln \mathcal{L}\left(n^{\rm IC},n^{\rm A}|\Phi^0_{\rm astro},\gamma_{\rm astro}\right) =\ln ( \mathcal{L^{\rm A}}\cdot  \mathcal{L^{ \rm IC}}) \,.
\label{eq:like}
\end{equation}
We observe that $\mathcal{L}$ is a function of the astrophysical flux normalization $\Phi_{\rm astro}^0$ and of the spectral index $\gamma_{\rm astro}$. The likelihood $\mathcal{L}$ is maximized with a statistical approach giving the best values for the two free parameters. The results are shown in Figures~\ref{fig:best} and~\ref{fig:bestpar}.

Figure \ref{fig:best} shows the neutrino spectrum for ANTARES (upper panel) and IceCube (lower panel). In the plots, the gray lines correspond to the sum of the best-fit for the neutrino signal and of the conventional atmospheric background (blue regions). The shaded regions represent the lower cuts in energy considered in the fit. Moreover, in case of ANTARES we are forced to consider an upper cut in energy ($E_\nu \leq300$~TeV) according to~\citet{proceedingA}, as the deduced ANTARES effective area is only known up to such an energy.

In Figure \ref{fig:bestpar} we show the contour plots for the fit of IceCube six-year HESE (black) and for the combined fit IceCube+ANTARES (ocher). The solid (dotted) lines correspond to the 68\% (95\%) confidence interval contours. The best-fit values and the 1-2$\sigma$ ranges of the flux normalization and the spectral index are obtained by marginalizing the two-dimensional likelihood and are reported in Table~\ref{tab}.
\begin{deluxetable}{|c|c|c|c|c|}
\tablecaption{Fitted parameters for the single power-law flux \label{tab}}
\tablehead{\colhead{Fit}&\colhead{Parameter}&\colhead{Best-fit}&\colhead{68\% C.I.}&\colhead{95\% C.I.}}
\startdata
IC & $\Phi_{\rm astro}^0$ &2.44&2.00 -- 2.94&1.62 -- 3.48 \\
 & $\gamma_{\rm astro}$ &2.95&2.76 -- 3.21&2.56 -- 3.46  \\ \hline
IC+A & $\Phi_{\rm astro}^0$ &2.30&1.90 -- 2.71&1.56 -- 3.16 \\
 & $\gamma_{\rm astro}$ &2.85&2.68 -- 3.04&2.52 -- 3.23  \\
\enddata
\tablecomments{Best-fit values and 1--2$\sigma$ intervals of $\Phi_{\rm astro}^0$ (in units of $10^{-18} \left({\rm GeV\,cm^2\,s\,sr}\right)^{-1}$) and $\gamma_{\rm astro}$ for the analysis on IceCube six-year HESE data (IC) and the combined analysis IceCube+ANTARES (IC+A).}
\end{deluxetable}
We note that, in case of the fit performed with IceCube data only, the best-fit values for the spectral index and the flux normalization differ from the ones reported by the IceCube Collaboration~\citep{proceedingIC} only by $1\%$ and $0.8\%$, respectively. It is worth noting that  the combined fit provides slightly smaller values for the flux normalization and the spectral index.

\section{The low-energy excess}

\begin{figure}[t!]
\centering
\includegraphics[width=0.48\textwidth]{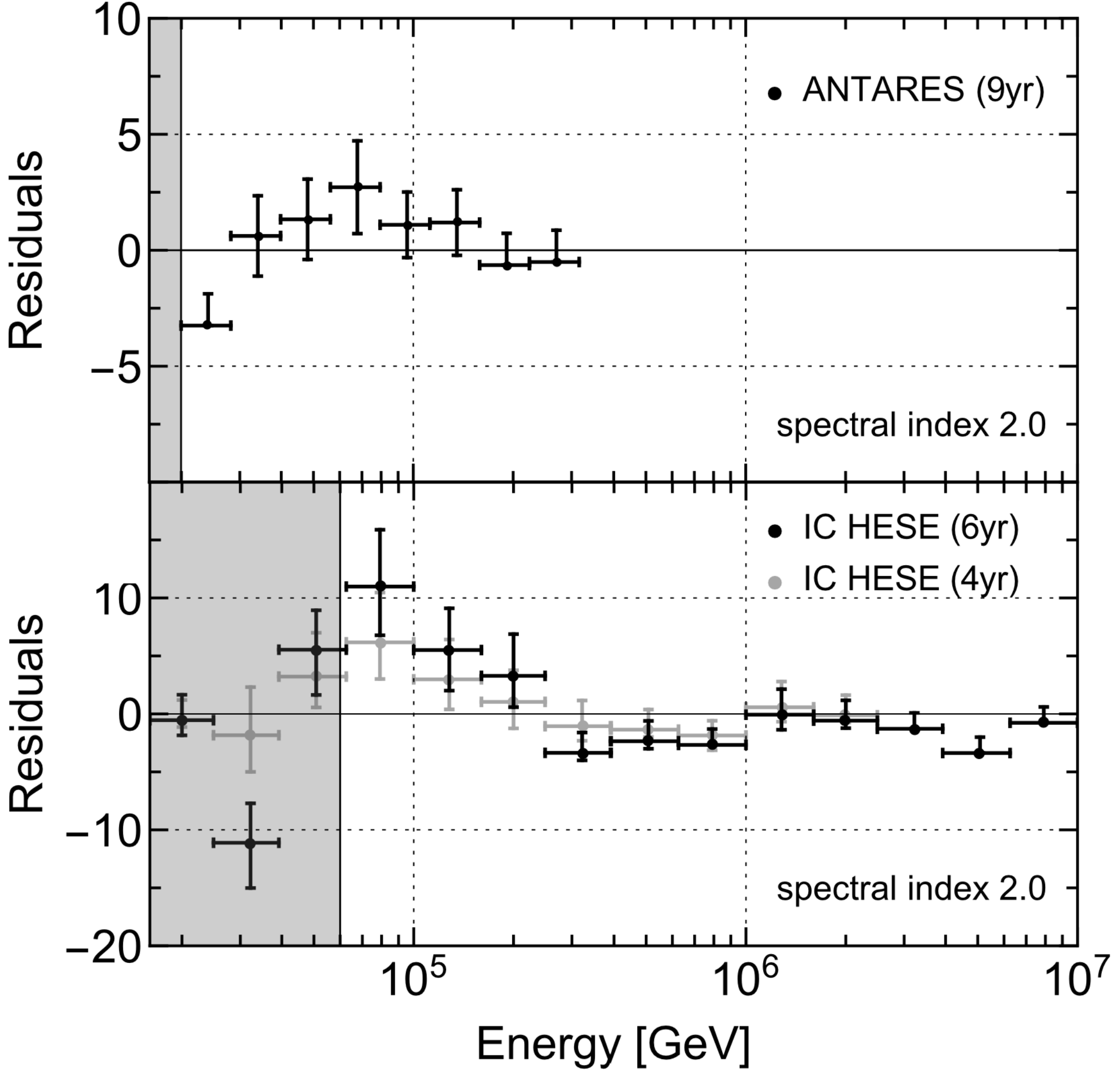}
\caption{\label{fig:residual}Residuals in the number of neutrino events as a function of the neutrino energy with respect to the sum of the conventional atmospheric background and a single astrophysical power-law with spectral index 2.0 for ANTARES (upper panel) and IceCube (lower panel). In the lower plot, we also report in gray the residuals corresponding to the IceCube four-year HESE, taken from~\citet{Chianese:2016opp}.}
\end{figure}

Finally, we study how the analysis changes if the spectral index is fixed to some specific values. For instance, a spectral index $\gamma_{\rm astro}=2.0$ is predicted by the standard Fermi acceleration mechanism, and is in general considered as a benchmark. In Figure~\ref{fig:residual} the residuals in the number of neutrino events for both experiments are reported once the sum of the conventional atmospheric background and of an astrophysical power-law $E_\nu^{-2.0}$ has been subtracted. The flux normalization is fitted by considering both IceCube and ANTARES data. Remarkably, both ANTARES and IceCube experiments seem to exhibit a slight excess in the same energy range (40--200~TeV). Moreover, we note that the local significance of the low-energy excess in IceCube increases if one considers six-year HESE data instead of four-year data.

The presence of an excess in both ANTARES and IceCube experiments has to be statistically tested. The null hypothesis is that in both samples the diffuse neutrino flux is just produced by an astrophysical power-law component superimposed to the conventional atmospheric background. In order to quantify the $p$-value for the null hypothesis, we perform a $\chi^{2}$ test. For Poisson-distributed data, the test statistics behaving as a $\chi^{2}$ with  $N - m$ dof is the following
\begin{equation}
\chi^{2}=-2\ln \mathcal{L}\left(n^{\rm IC},n^{\rm A}|\Phi^0_{\rm astro},\gamma_{\rm astro}\right)\,,
\label{eq:chi}
\end{equation}
where $\mathcal{L}\left(n^{\rm IC},n^{\rm A}|\Phi^0_{\rm astro},\gamma_{\rm astro}\right)$ is defined in Eq.~(\ref{eq:like}). Note that $N=18$ is the total number of energy bins and $m=1$ the number of free parameters in the fit. We perform the test for different values of the spectral index $\gamma_{\rm astro}$, while the flux normalization $\Phi^0_{\rm astro}$ is obtained by maximizing the likelihood.

Because the previous $\chi^{2}$ analysis has to be performed when the events of each bin are Gaussian distributed, a condition that in principle could not be satisfied for small number of events, we additionally perform a more general non-parametric test, namely the one-dimensional Kolmogorov-Smirnov~(KS) statistical test. For each experiment, the test compares the empirical cumulative distribution function deduced by data with the one obtained under the null hypothesis of power-law behavior. For a given spectral index, the $p$-value is evaluated by a bootstrap method for IceCube and ANTARES experiments, respectively. The two $p$-values are then combined by means of the Fisher's method. Note that, in order to perform the test, it would be necessary to know the list of ANTARES events with their measured energy, which is still not available. To avoid such a limitation, we reasonably assume a homogeneous distribution of the events in each energy bin of the ANTARES data set. We do not expect that a detailed knowledge of the events energy would dramatically change our results due to the large uncertainty on the energy measurement.
\begin{figure}[t!]
\centering
\includegraphics[width=0.48\textwidth]{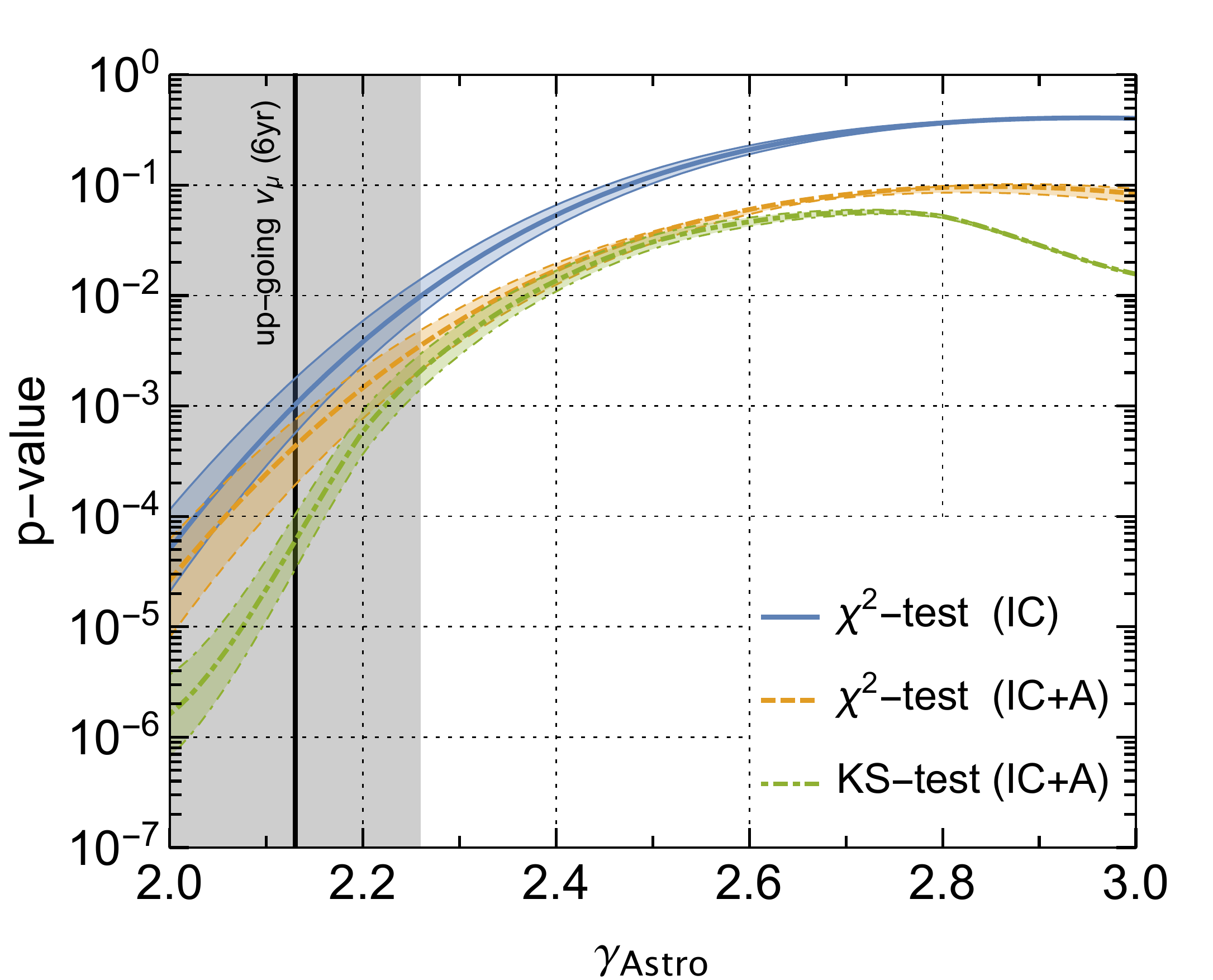}
\caption{\label{fig:pvalue}The solid (dashed) line represents the $p$-value as a function of the spectral index  for  the $\chi^{2}$-test for  the IC (IC+A) data sample. The dotted-dashed line refers to the KS test performed by combining the two data samples with the Fisher's method. The bands correspond to an uncertainty of $\pm 20\%$ on the conventional background estimation. The vertical band, instead, represents the best-fit for the spectral index as deduced by six-year up-going muon neutrinos $(\gamma=2.13\pm0.13$;~\citealt{Aartsen:2016xlq}).}
\end{figure}

Figure~\ref{fig:pvalue} shows the $p$-value for the $\chi^{2}$ and KS hypothesis tests as a function of the spectral index adopted in the analysis. The bands are obtained by considering an uncertainty of $\pm 20\%$ on the conventional background estimation in both experiments. As one can see from the plot, the addition of the ANTARES data set has the effect of reducing the $p$-value (by about a factor 2$\div$3) independently of the assumed spectral index. This means that fixing a certain threshold in $p$-value for rejecting the null hypothesis, the addition of the ANTARES data set to the IceCube one enlarges the range of spectral indexes for which the null hypothesis is disfavored. Moreover, we observe that the interpretation of the six-year up-going muon neutrinos as a single power-law with $\gamma=2.13\pm0.13$~\citep{Aartsen:2016xlq} is almost statistically incompatible ($p$-value smaller than $10^{-2}$) with the same interpretation for the whole data sample. We underline that the benchmark prediction of Fermi acceleration mechanism $\gamma=2.0$ has a $p$-value equal to $2.6^{+3.6}_{-1.8}\times10^{-5}$ for $\chi^{2}$ and $1.6^{+2.1}_{-1.0}\times10^{-6}$ for KS statistical tests, where the errors correspond to a $\pm20\%$ uncertainty on the conventional atmospheric background. This means that the theoretical prior of a single astrophysical power-law component with a spectral index that is dictated by acceleration mechanisms {\it \`a la} Fermi is more disfavored by the combined data sets with respect to the stand-alone IceCube sample.

\section{Conclusions}

In summary, we have scrutinized the hypothesis that a single astrophysical power-law component describes the diffuse neutrino flux observed by both ANTARES (nine-year showers) and IceCube (six-year showers and tracks) experiments. The combined analysis of the ANTARES and IceCube preliminary data samples gives values for the flux normalization and the spectral index best-fit that are 3\% and 6\%, respectively, lower than those obtained with IceCube alone. Interestingly, both data samples show an excess in the same energy range (40--200~TeV) that could suggest the presence of an additional contribution. In order to quantify such a hint, we have performed a goodness-of-fit test on the null hypothesis of a single power-law, scanning different values for the spectral index. The addition of the ANTARES data set generally reduced the $p$-value by about a factor 2$\div$3. Moreover, a single power-law component in the neutrino flux with the spectral index deduced by the six-year up-going muon neutrinos of IceCube is disfavored with a $p$-value smaller than $10^{-2}$. Such an analysis shows once more how the synergy among different NTs can improve and reinforce the results of a single experiment, and this will be even more true for future large NT such as KM3NeT~\citep{Adrian-Martinez:2016fdl} and IceCube-Gen2~\citep{Aartsen:2014njl}.
 
\acknowledgments
We thank Luigi~Antonio~Fusco for the useful discussion.

\end{document}